\documentclass[pra,twocolumn]{revtex4}
\usepackage[english]{babel}
\usepackage{amsmath}
\usepackage{amssymb}
\usepackage{epsfig}

\begin{document}

\title{Bubbles with attached quantum vortices in trapped binary Bose-Einstein condensates}
\author{Victor P. Ruban}
\email{ruban@itp.ac.ru}
\affiliation{Landau Institute for Theoretical Physics, RAS,
Chernogolovka, Moscow region, 142432 Russia} 
\date{\today}

\begin{abstract}
Specific topological excitations of energetically stable ``core-and-mantle'' configurations
of trapped two-component immiscible Bose-Einstein condensates are studied numerically 
within the coupled Gross-Pitaevskii equations. Non-stationary long-lived coherent structures,
that consist of several quantum vortex filaments penetrating the ``mantle'' from outside to
inside and vice-versa and demonstrate quite nontrivial dynamics, are observed in simulations 
for the first time. The ends of filaments can remain attached to the interface between the
``mantle'' and the ``core'' if the latter is large enough while the surface tension is not 
small. The shapes of such ``bubbles'' are strongly affected by the vortices and sometimes 
are far from being spherical. 
\end{abstract}

\maketitle
%%%%%%%%%%%%%%%%%%%%%%%%%%%%%%%%%%%%%%%%%%%%%%%%%%%%%%%%%%%%%%%%%%%%%%%%%%%%%%%%

\section{Introduction}

Multi-component mixtures of ultracold Bose-Einstein-condensed atomic gases have been 
extensively studied over a quarter of century \cite{mix1,mix2,mix3,mix4,mix5}. 
Such systems consist either of different chemical elements, or of different isotopes
of the same element, or of the same isotope in different internal (hyperfine) quantum 
states. The interactions between different species give rise to a rich variety of unique 
features which do not exist for a single-component Bose-Einstein condensate (BEC). 
What is very important, parameters of nonlinear interactions for matter waves, being
proportional to the scattering lengths, can be tuned in many cases via Feshbach 
resonances \cite{mix-Rb85-87-tun,tun1,tun2,tun3,RMP2010}. In particular, 
with a sufficiently strong cross-repulsion between two components, phase separation 
occurs \cite{separation,AC1998}. It lies in the base of many related configurations 
and phenomena of high interest, such as domain walls and surface tension between 
segregated condensates \cite{mix4,tension}, nontrivial ground state geometry of 
binary immiscible condensates in traps \cite{gr1,gr2,gr3} (including optical lattices
\cite{latt1,latt2,latt3}), the dynamics of bubbles \cite{bubbles}, 
the quantum counterparts of classical hydrodynamic instabilities 
(Kelvin-Helmholtz \cite{KHI1,KHI2}, Rayleigh-Taylor \cite{RTI1,RTI2,RTI3},
Plateau-Rayleigh \cite{capillary}, the parametric instability of capillary waves at 
the interface \cite{param_inst-1,param_inst-2}), complex textures in rotating binary 
condensates \cite{mix-sheet-1,mix-sheet-2,topo_defects}, vortices with filled cores
\cite{mix3,VB1,VB2,VB3,massive-vort-2D-1,massive-vort-2D-2,R2021-1},
three-dimensional topological structures
\cite{vortex-mol,wall-annih-1,wall-annih-2,vortex-wall,handles}, 
capillary flotation of dense drops in trapped immiscible BECs \cite{R2021-2}, etc.

Among other coherent structures, quantum vortices have long been recognized as objects
of primary interest and importance. Already for single-component condensates, a lot of
expressive results has been obtained on vortex configurations and their dynamics 
(see, e.g., \cite{BEC1,BEC2,F2009,SF2000,R2001,AD2003,ring_inst,reconn-2017,TWK2018,
R2018,TRK2019,shell-BEC}, 
and references therein). Concerning vortices in mixed condensates, the realm is even
larger and containing wide unknown territories. Many ``godsends'' are still possible 
there for an explorer. This work presents a new kind of rather elegant long-lived 
compound vortex-bubble structures with nontrivial dynamics. To explain what they 
basically are, let us recall that in a two-component immiscible condensate, domain walls 
with attached quantum vortices are possible \cite{vortex-wall}. They are highly deformed 
vortex sheets of a special form. In Ref.\cite{vortex-wall}, the authors investigated such
complexes for the case of equal self-repulsion coefficients. Numerical solutions were
obtained there for binary BECs in elongated traps where the equilibrium interface between 
the species was a disc with the edge at the Thomas-Fermi surface. Quantum vortices were 
present in both components and directed mainly along the trap axis.

Here it will be shown that essentially new and interesting feature arises when there 
is inequality (asymmetry) between the coefficients. In a nearly spherical harmonic trap,
it results in formation of a stable and compact background equilibrium configuration of 
the ``core-and-mantle'' type. The presence of stably trapped, vortex-free ``core'' at the 
center, in combination with the surface tension between the components, tends to stabilize
possible vortex filaments attached to such a bubble from outside and penetrating the 
``mantle'' (see Fig.1 for example). Of course, the number of (outgoing) vortices is always 
equal to the number of (ingoing) antivortices. The density of the outer component is so 
negligible inside the ``core'' that practically it does not matter which of the vortices 
is the continuation of a given antivortex. Only the overall balance is usually essential.
Each vortex or antivortex typically retains its identity and is directed roughly along the 
local radius. Due to interactions, the vortices are in transversal motion, simple or complicated.
But sometimes, a pair of attached filaments (one ingoing and one outgoing) can dynamically 
couple, then get detached from the bubble and finally form a separate vortex filament. 
However, in such scenario the vortex pair has to overcome a ``potential barrier'' caused 
by surface tension. For relatively large bubbles and a small number of vortices, such 
processes occur quite rarely. Therefore structures of this kind are able to exist for 
thousands trap time units, as our numerical simulations show. We will also see that the 
dynamics becomes progressively more complicated with increasing number of attached vortices.

\begin{figure}
\begin{center}
\epsfig{file=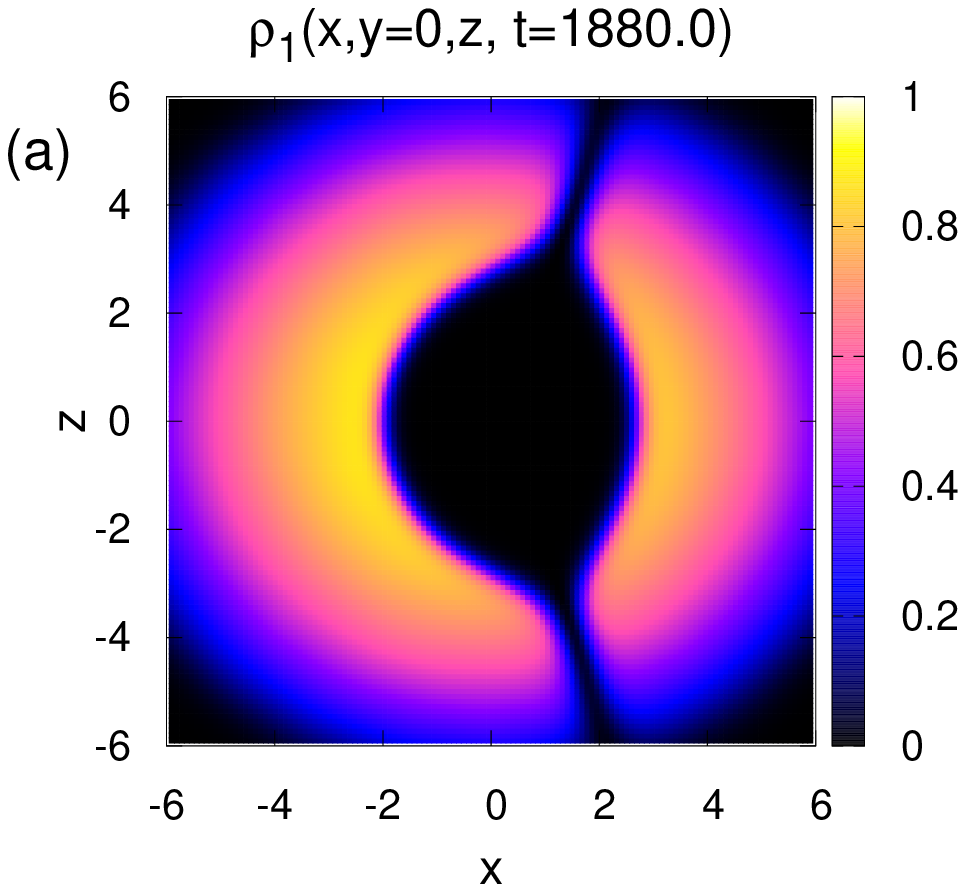, width=42mm}
\epsfig{file=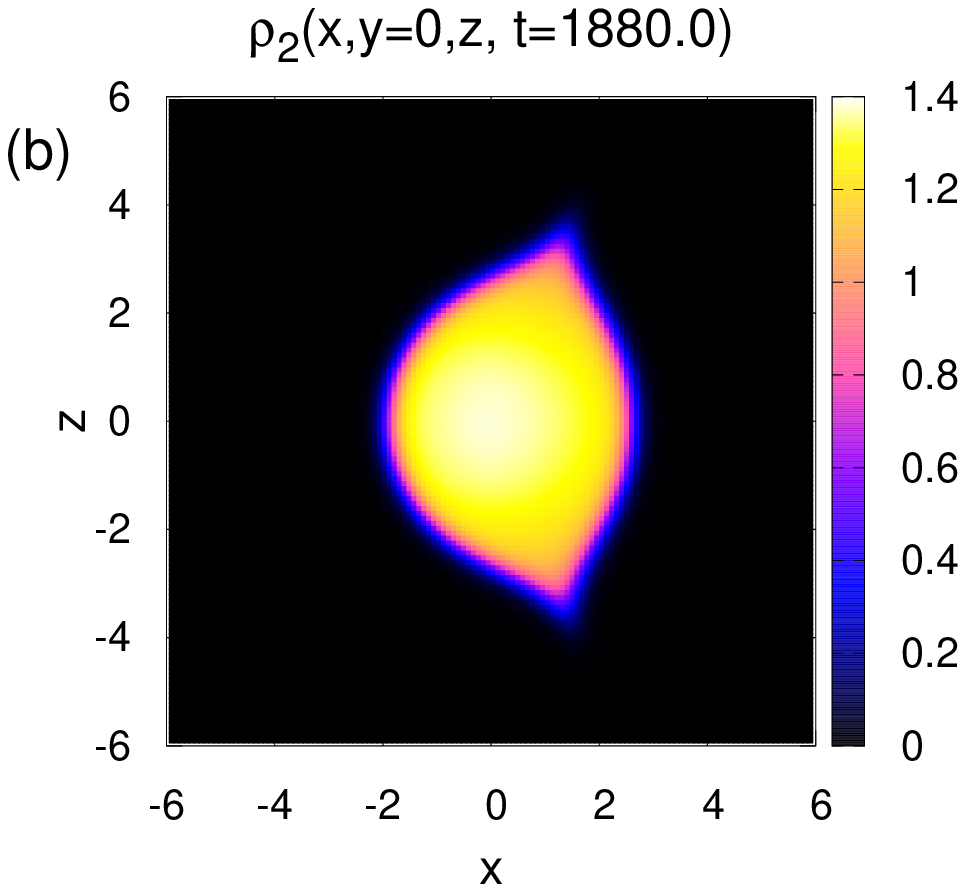, width=42mm}
\end{center}
\caption{
A numerical example of long-lived trapped condensate bubble with one pair of 
attached vortex filaments. Presented are normalized densities of the two condensate 
components in cross-section $y=0$ (note the difference in the color scales): 
a) ``mantle''; b) ``core''. Transition layer between the clouds is rather sharp.
The configuration is slowly rotating around $z$ axis, with super-imposed oscillations
of the interface, and it is shown at the moment when both vortices lie in plane $y=0$. 
In this simulation $\lambda=1.1$, $g_{11}=1.0$, $g_{22}=0.6$, $g_{12}=1.2$, 
$n_1=1521.8$, $n_2=175.8$, and $\mu_1=30$.}
\label{V1_example} 
\end{figure}

To the best of my knowledge, such three-dimensional complexes in trapped immiscible BECs 
have not yet been discussed in the literature. Far-distant analogies are however known, 
as quantum vortices in crusts of neutron stars (see, e.g., review \cite{ns}), 
or ${}^3$He-${}^4$He topological structures suggested  in Ref.\cite{V2000} (a droplet of 
${}^4$He immersed in the ${}^3$He liquid). The purpose of this work is to introduce bubbles 
with attached quantum vortices theoretically in the context of ultracold gases and illustrate, 
on several representative numerical examples, their main properties. Scientific importance of
these new structures is in their theoretical existence (and hopefully experimental feasibility) 
within wide and realistic parameter ranges, together with their quite nontrivial dynamics.

\section{The model and numerical method}

As the basic mathematical model in our research, we employ the widely recognized coupled 
Gross-Pitaevskii equations for two complex wave functions, $A({\bf r},t)$ 
(the first component, ``mantle''), and $B({\bf r},t)$ (the second component, ``core''). 
This conservative model is applicable for rarefied Bose-gases in the limit of zero temperature.
For simplicity, equal masses $m_1=m_2=m$ of the species atoms are considered 
(or the small difference in mass of isotopes as, for example, $^{85}$Rb and $^{87}$Rb 
is neglected \cite{tun2}). Let an axisymmetric harmonic trap be characterized by 
a perpendicular frequency $\omega_\perp$ and by an anisotropy 
$\lambda=\omega_\parallel/\omega_\perp$. With the trap units $\tau=1/\omega_\perp$ 
for the time, $l_{\rm tr}=\sqrt{\hbar/\omega_\perp m}$ for the length, and  
$\varepsilon=\hbar\omega_\perp$ for the energy,
the equations of motion are written in dimensionless form 
\begin{eqnarray}
i\dot A=-\frac{1}{2}\nabla^2 A+\left[V(x,y,z)+g_{11}|A|^2+g_{12}|B|^2\right]A,&&
\label{GP1}\\
i\dot B=-\frac{1}{2}\nabla^2 B+\left[V(x,y,z)+g_{21}|A|^2+g_{22}|B|^2\right]B,&&
\label{GP2}
\end{eqnarray}
where  
$V=(x^2+y^2+\lambda^2 z^2)/2$ is the trap potential, and $g_{\alpha\beta}$ 
is the symmetric $2\times 2$ matrix of nonlinear interactions. Physically, 
the interactions are determined by the scattering lengths $a_{\alpha\beta}$ \cite{mix2}:
\begin{equation}
g_{\alpha\beta}^{\rm phys}=2\pi \hbar^2a_{\alpha\beta}(m_\alpha^{-1}+m_\beta^{-1}).
\end{equation}
We are interested in the case of all positive $g_{\alpha\beta}$. Without loss of generality, 
the first self-repulsion coefficient can be normalized to the unit value, $g_{11}=1$, 
since we consider $g_{\alpha\beta}$ as constant in time parameters throughout this work.
With this choice, the (conserved) numbers of trapped atoms are given by relations
\begin{eqnarray}
&&N_1=\frac{l_{\rm tr}}{4\pi a_{11}}\int |A|^2 d^3{\bf r}=(l_{\rm tr}/a_{11}) n_1,\\
&&N_2=\frac{l_{\rm tr}}{4\pi a_{11}}\int |B|^2 d^3{\bf r}=(l_{\rm tr}/a_{11}) n_2.
\end{eqnarray} 
In real experiments the ratio $l_{\rm tr}/a_{11}$ ranges typically from a few hundreds
to a few thousands.

It is a well known fact that system (\ref{GP1})-(\ref{GP2}) in the phase-separated 
regime is similar to potential flows in the classical hydrodynamics of two immiscible 
compressible fluids, except for the vicinity of domain walls and vortex cores where 
the ``quantum pressures'' come into play.
The ``hydrodynamic pressures'' of the first and the second matter-wave ``fluids'' are 
$g_{11}|A|^4/2$ and $g_{22}|B|^4/2$ respectively, and across the interface they are 
approximately equal (the difference caused by surface tension and by slow flows is 
relatively small). At equal pressures, the component with smaller self-repulsion 
coefficient is more dense. Therefore, in order to ensure a relatively dense and stable 
core, it is necessary to satisfy the inequality $g_{22}<g_{11}=1$. 

It is a known fact that equilibrium states are characterized by two chemical potentials 
$\mu_1$ and $\mu_2$. Under relevant conditions $\mu_1\gg 1$ and $\mu_2\gg 1$, 
the background density profiles are given by the Thomas-Fermi approximation:
\begin{eqnarray}
&&|A_0|^2\approx [\mu_1-V(x,y,z)]\equiv\mu_1 \rho_{1 \rm{eq}}({\bf r}),
\label{AA_eq}\\
&&|B_0|^2\approx [\mu_2-V(x,y,z)]/g_{22}\equiv\mu_1 \rho_{2 \rm{eq}}({\bf r}),
\label{BB_eq}
\end{eqnarray} 
where we have introduced the normalized densities $\rho_1=|A|^2/\mu_1$ and 
$\rho_2=|B|^2/\mu_1$. The first formula is valid in the mantle, while the second one 
in the core. An effective transverse size of the condensate is thus $R_\perp=\sqrt{2\mu_1}$,
while a characteristic width of vortex lines in the mantle is $\xi\sim 1/\sqrt{\mu_1}$.

The following condition for phase separation is assumed \cite{separation,AC1998},
$$
g=(g^2_{12}-g_{11}g_{22})>0. 
$$
There is a narrow transition layer between the segregated condensates
(its width is roughly proportional to $1/\sqrt{g}$; it is also seen in Fig.1) 
and the associated surface tension roughly proportional to $\sqrt{g}$ 
\cite{separation,tension}. Numerical simulations performed here were oriented 
on experimentally realizable $^{85}$Rb-$^{87}$Rb mixtures \cite{tun2},
where $a_{12}/a_{22}\approx 2$ while $a_{11}$ can be tuned via Feshbach resonance. 
Therefore in all our runs $g_{12}=2g_{22}$. On the other hand, an optimal value 
for the ratio $g_{22}/g_{11}$ should simultaneously provide a large surface tension, 
and be not close to unity for core stability. We typically put $g_{22}=0.5$ or $0.6$ 
in this work, but in some runs, for comparison, $g_{22}=0.3$ or $0.8$.

The coupled Gross-Pitaevskii equations (\ref{GP1})-(\ref{GP2}) were simulated 
using a standard Split-Step Fourier Method of the second order accuracy. Spatial 
and temporal resolutions were sufficient to conserve the corresponding Hamiltonian 
functional and the numbers of particles $n_1$ and $n_2$ up to the 5th decimal place
over time period $T_{\rm run}=2000$. 

Initial states containing only soft excitations were prepared by application of 
the gradient descent procedure, which in the given case is equivalent to an 
imaginary-time propagation for a finite period of an auxiliary time-like variable. 
The input for this dissipative procedure was a configuration of the first component 
as $A_{\rm inp}=\tilde A_0({\bf r})C_0({\bf r})$, with a real $\tilde A_0({\bf r})$.
Several vortex filaments were introduced through the complex multiplier
\begin{equation}
C_0({\bf r})=\prod_{j} \frac{w_j}{\sqrt{|w_j|^2+\epsilon}}, 
\end{equation}
where explicit form for $w_j({\bf r})$ depended upon the desired input vortex shape. 
In particular, a filament oriented roughly in $z$ direction was introduced by 
\begin{equation}
w_j=[x-X_j(z)]\pm i[y-Y_j(z)],
\end{equation}
with relatively small arbitrary functions $X_j(z)$ and $Y_j(z)$. Accordingly, $x$ or $y$ 
variable parameterized vortex filaments oriented closely to $x$ or $y$ direction,
respectively. A small positive $\epsilon$ was employed to avoid zero in the denominators. 
The choice $\pm$ imposed the sign of rotation of the $j$-th vortex. In the output, every 
such filament gave one vortex and one antivortex in the above discussed sense. The ``seed''
$B_{\rm inp}=\tilde B_0({\bf r})$ for the second component contained no vortices. 

The dissipative procedure ``filtered out'' fast potential excitations
and brought the system density closely to its background equilibrium configuration 
Eqs.(\ref{AA_eq}-\ref{BB_eq}), including the specific narrow wall profile between 
the mantle and the core. Correct cross-sections of the vortices were formed as well,
adjusted to the background density. 
At the same time, slow excitations as vortex positions could not achieve a minimum 
of their energy during a relatively short imaginary-time propagation,
and this fact resulted in subsequent nontrivial vortex dynamics at the main,
conservative stage of simulation. Taking different values for $\mu_1$ and $\mu_2$, 
various vortex configurations $X_j(z)$ and $Y_j(z)$ as the input of the dissipative 
procedure, and different periods for the imaginary-time propagation, 
it was possible to achieve different values of $n_1$, $n_2$, as well as various
initial arrangements of the attached vortices.

The trap anisotropy was chosen not far from unity, typically $\lambda=1.1$.
That was done in order to break the spherical symmetry, but just slightly.

It should be said that no short-scale instabilities were observed in our simulations.
The structures never decay if a small perturbation is present.
This fact can be viewed as an indirect evidence of their dynamical stability.
Some of the obtained numerical results are presented and discussed in the next section.

\section{Results}

\begin{figure}
\begin{center}
\epsfig{file=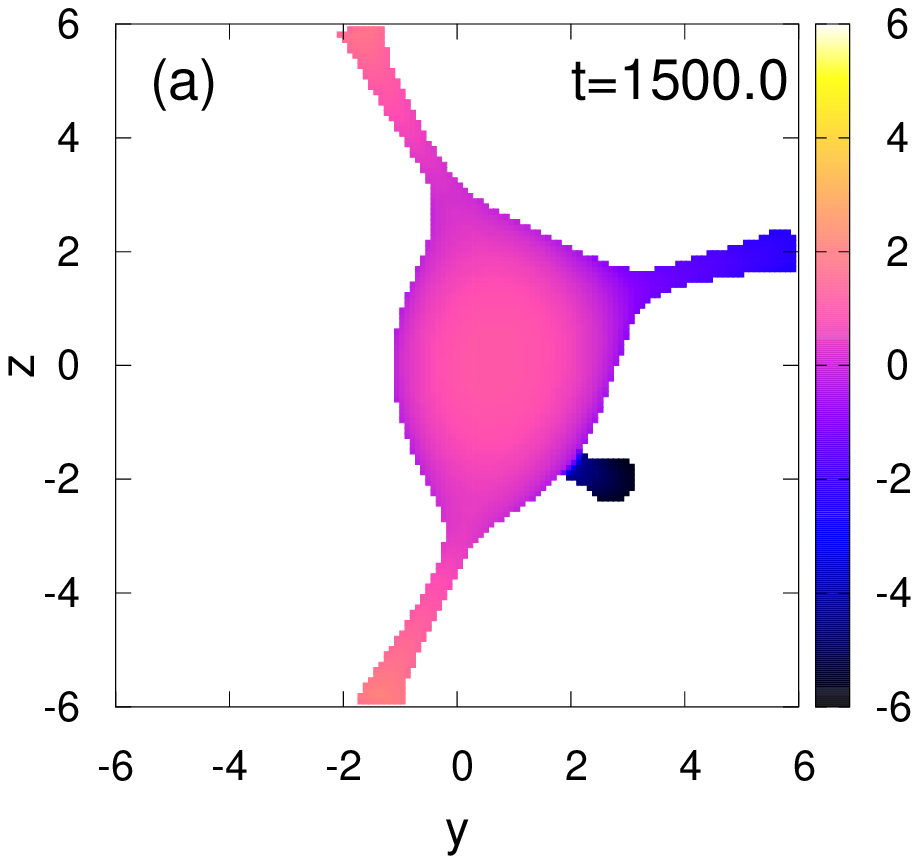, width=42mm}
\epsfig{file=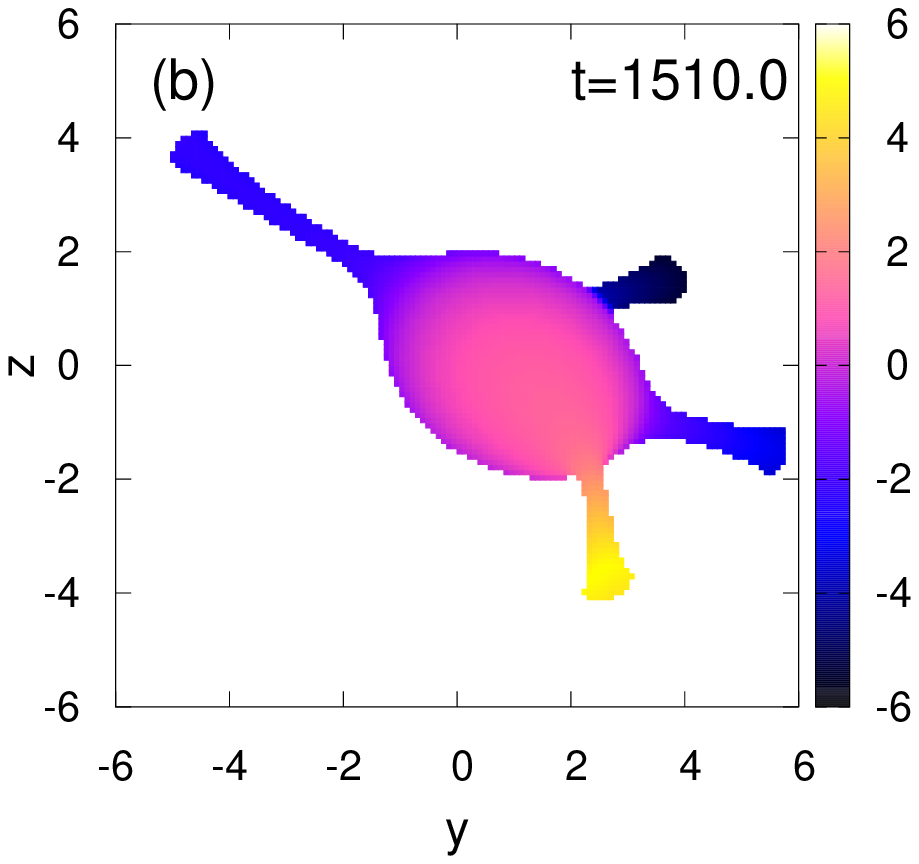, width=42mm}\\
\epsfig{file=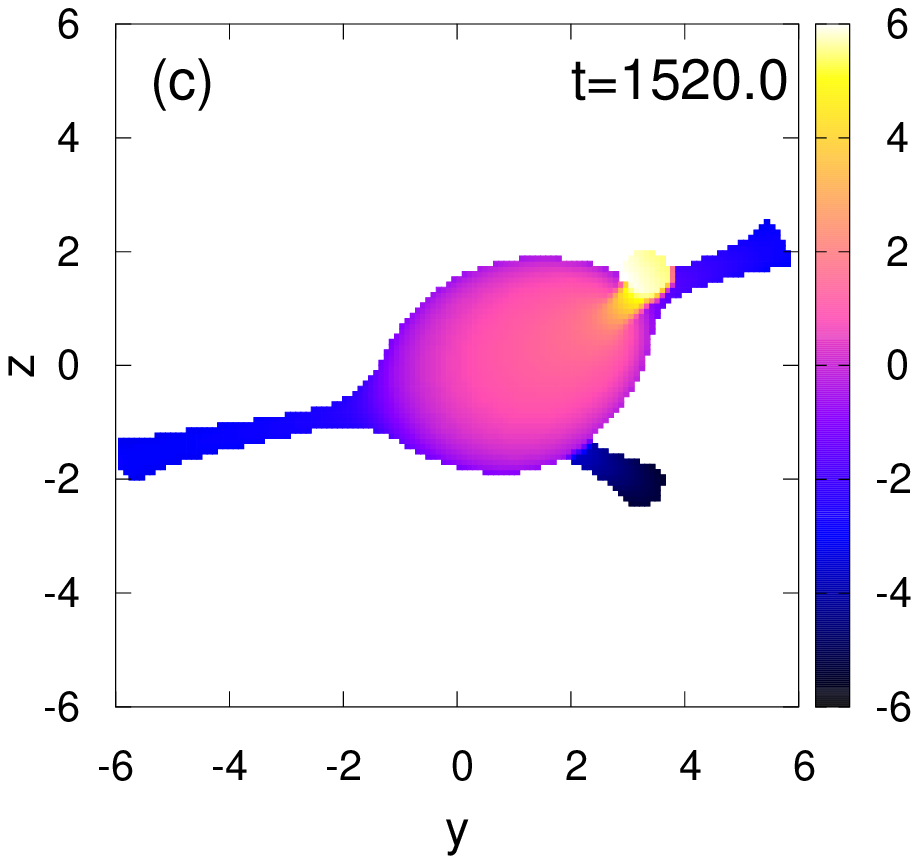, width=42mm}
\epsfig{file=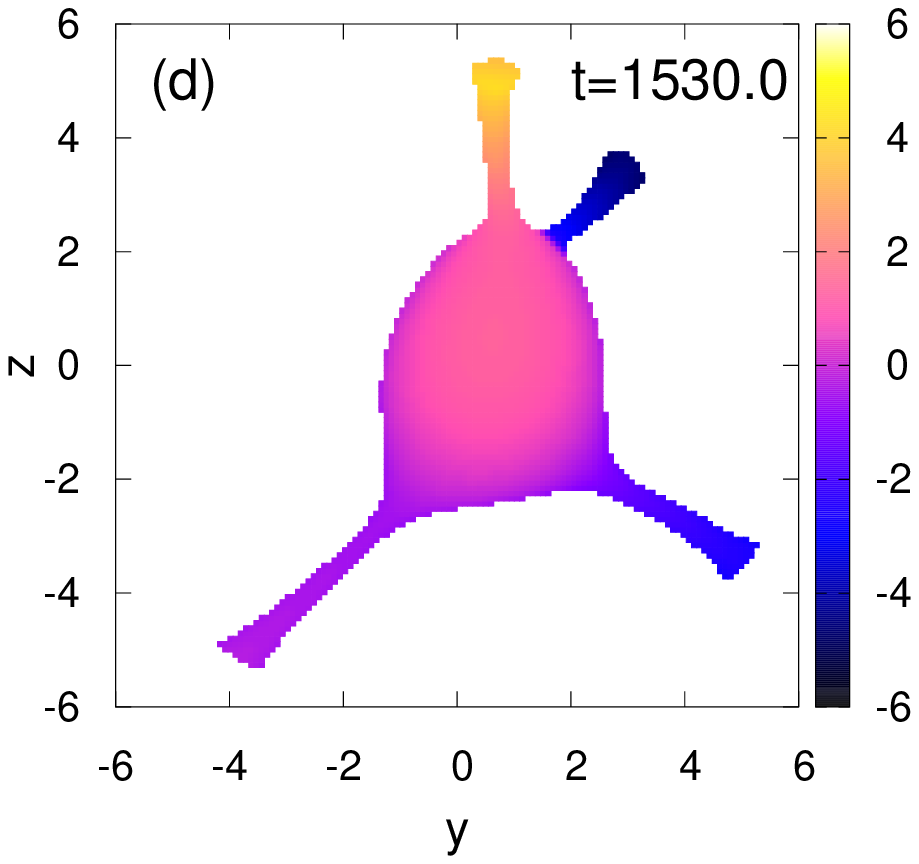, width=42mm}
\end{center}
\caption{A long-lived nonstationary bubble with two pairs of attached vortices. 
The color scale corresponds to $x$-coordinate of the grid points nearest to the 
surface determined by equation $\rho_1({\bf r},t)= 0.5\rho_{1 \rm{eq}}({\bf r})$. 
Only points within the domain $\rho_{1 \rm{eq}}({\bf r})>0.2$ are shown, 
and therefore the outer ends of filaments are truncated. In this simulation 
the structure is not simply rotating around a time-dependent instant axis 
(not aligned with $z$-axis), but also is subjected to slow deformations.
In general, it resembles somersaults. The parameters are:  $\lambda=1.1$, 
$g_{11}=1.0$, $g_{22}=0.6$, $g_{12}=1.2$, $n_1=1552.2$, $n_2=117.9$, and $\mu_1=30$.}
\label{V2_example} 
\end{figure}

The simplest case of trapped bubble with just one pair of attached vortex filaments 
is presented in Fig.1. It is clearly seen there that the ``kernel'' is far from being 
spherical, since the vortex ends pull the interface. In particular, the kernel boundary
has sharp conical features near the vortex ends. In this example, the initial vortex 
configuration was nearly symmetric with respect to the equatorial plane $z=0$, 
so the motion is a simple slow rotation around $z$-axis accompanied by bubble
oscillations due to ``imperfect'' initial conditions (see video \cite{video1}). 
In another run, without equatorial symmetry, the motion was slightly less monotonous, 
because an instant axis of rotation depended on time (not shown).

The behavior becomes nontrivial starting from two pairs of attached vortices. An example 
is shown in Fig.2 where the shape of an effective (inner) boundary of the first component
is presented for several time moments. In this case, the motion resembles some gymnastic 
performance, with somersaults, or a skydiver tumbling through the air 
(see video \cite{video2}). Of course, more ``tranquil'' regimes are also possible, 
when a nearly stationary configuration is rotating around $z$-axis, similarly to the 
previous example.

\begin{figure}
\begin{center}
\epsfig{file=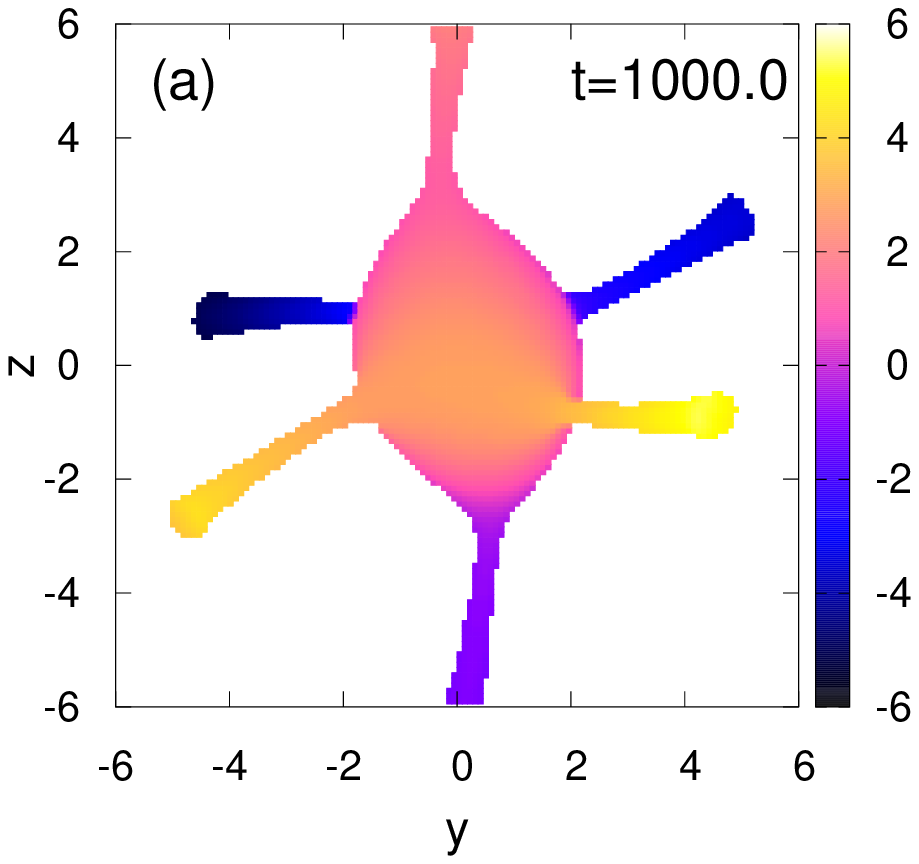, width=42mm}
\epsfig{file=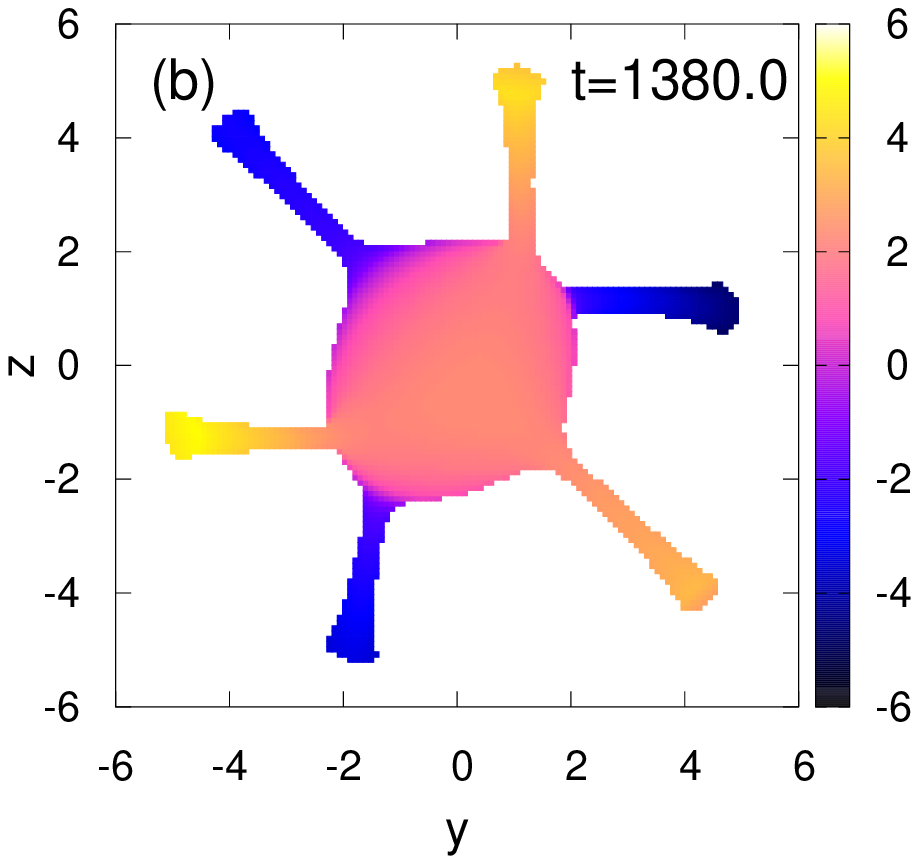, width=42mm}\\
\epsfig{file=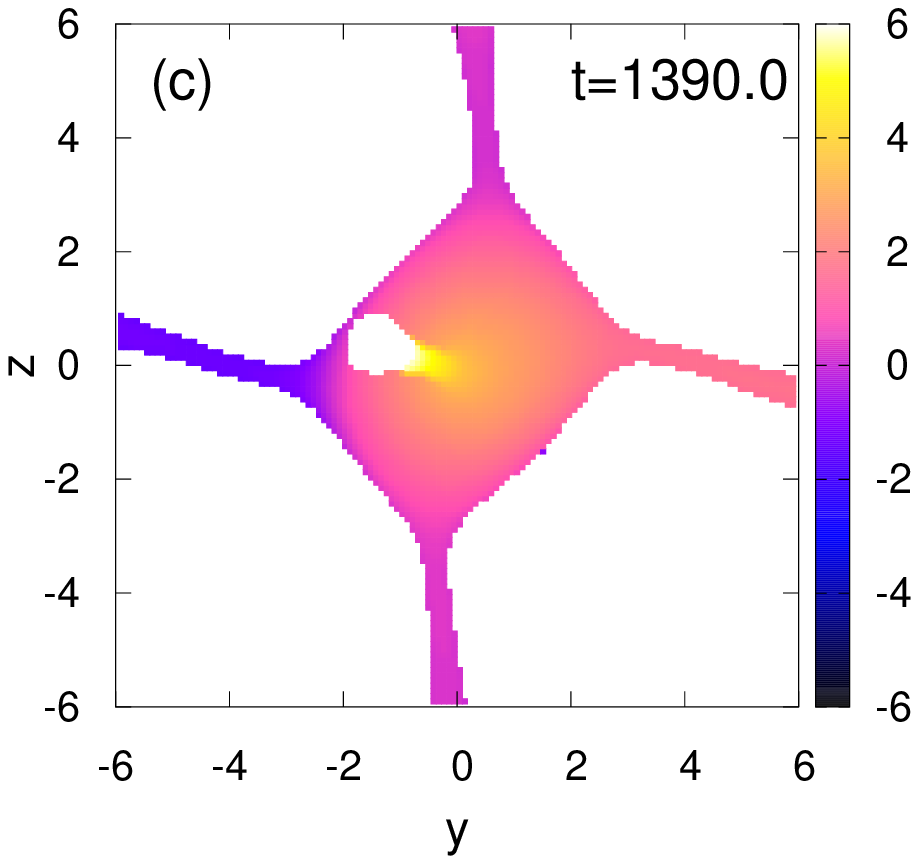, width=42mm}
\epsfig{file=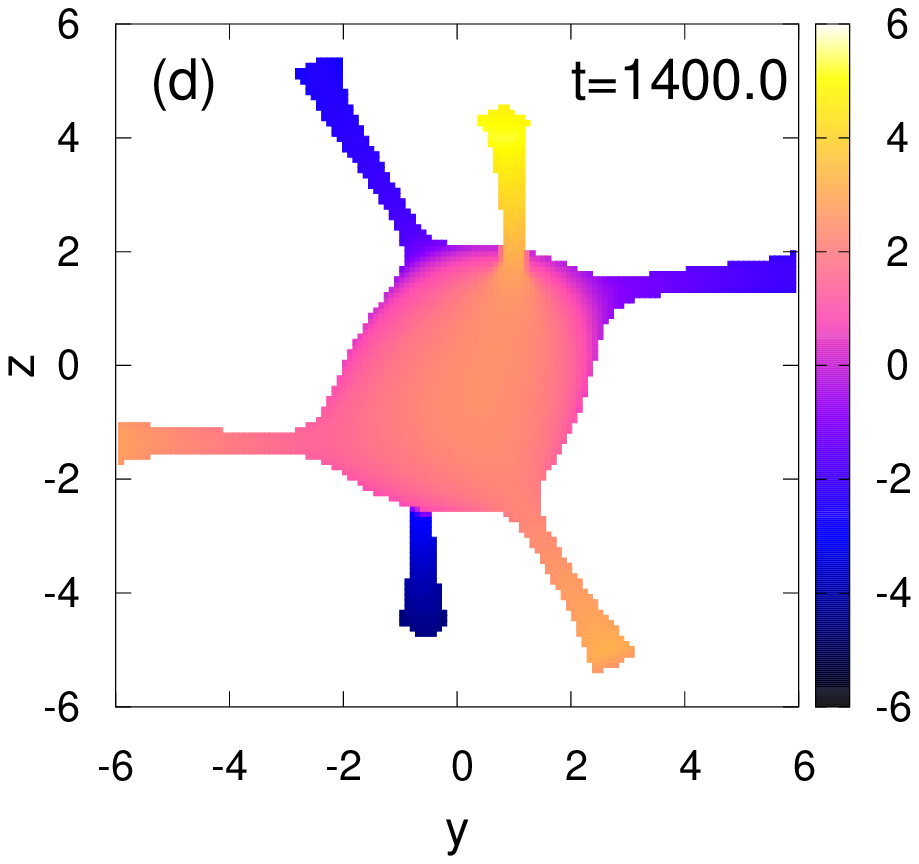, width=42mm}
\end{center}
\caption{A regularly rotating bubble with three pairs of attached vortices. 
The parameters are:  $\lambda=1.1$, 
$g_{11}=1.0$, $g_{22}=0.8$, $g_{12}=1.6$, $n_1=1503.3$, $n_2=141.7$, and $\mu_1=30$.
Initially, the vortices were oriented roughly along the three Cartesian axes.
The instant rotational axis is slowly precessing around $z$-direction.}
\label{V3_example1} 
\end{figure}

\begin{figure}
\begin{center}
\epsfig{file=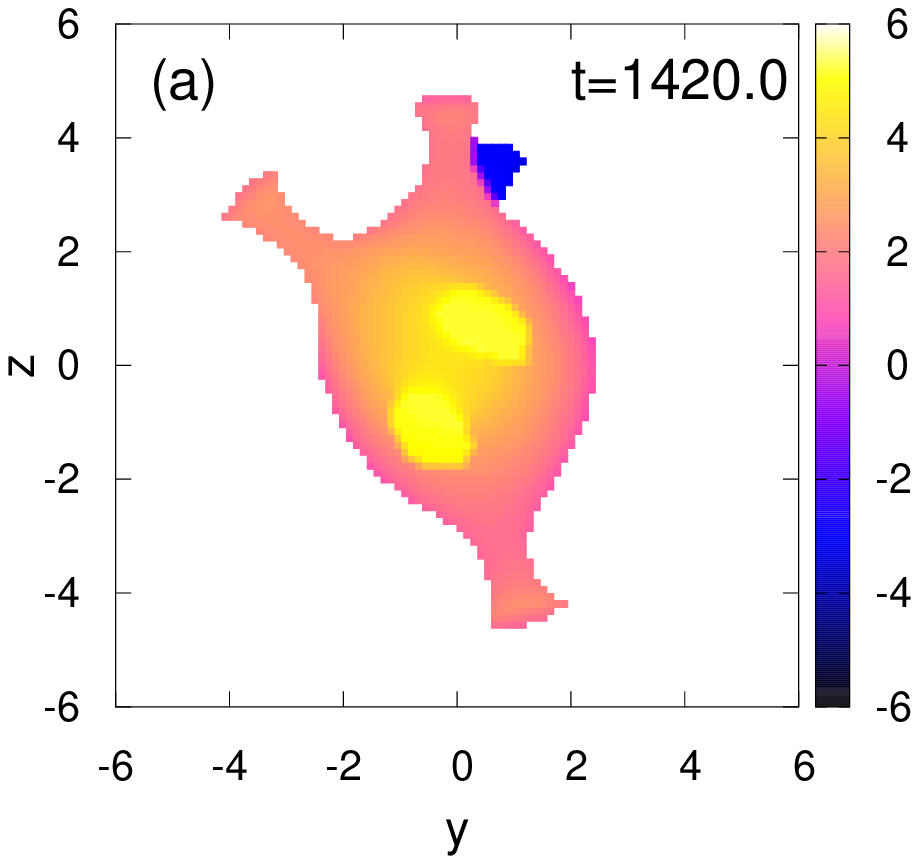, width=42mm}
\epsfig{file=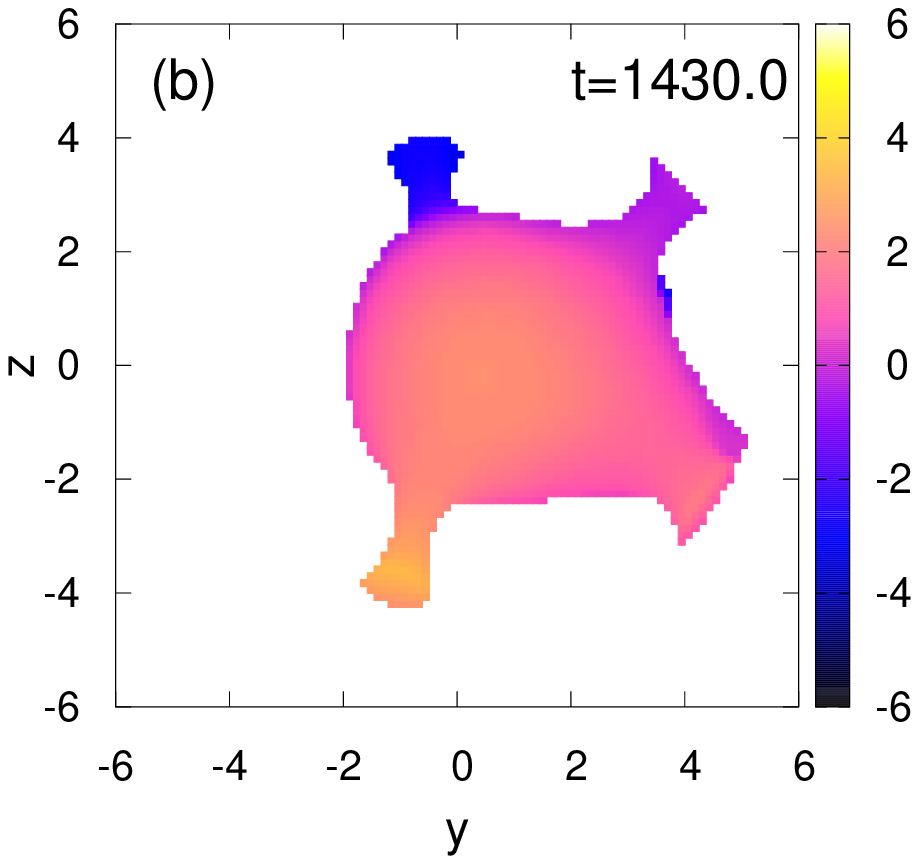, width=42mm}\\
\epsfig{file=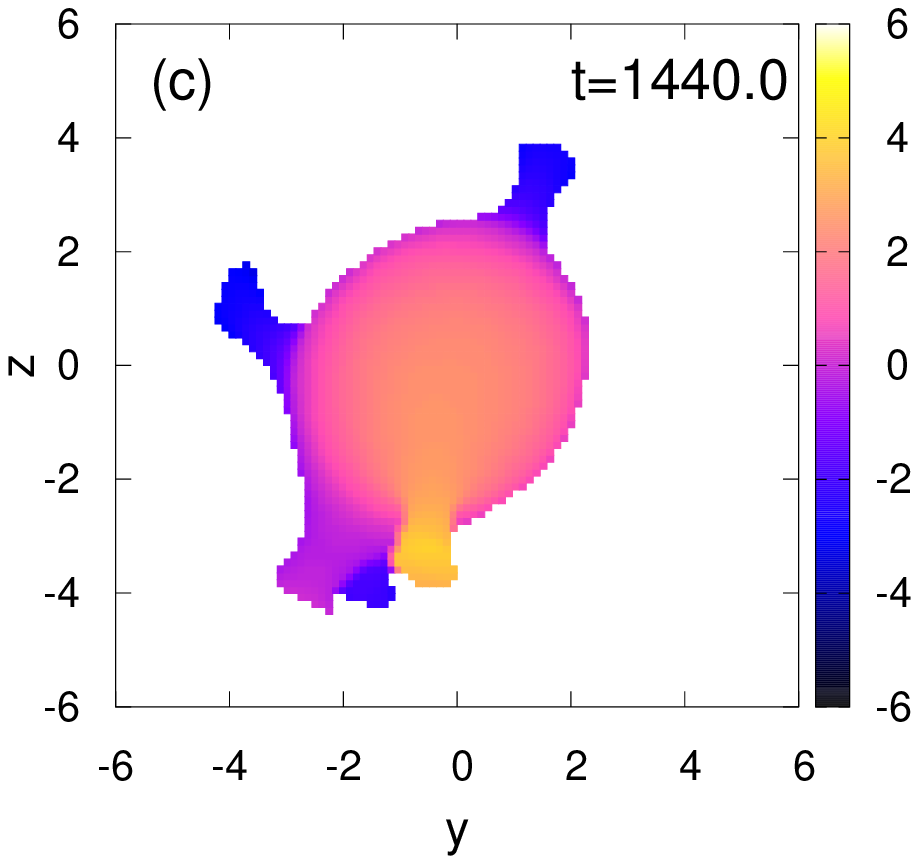, width=42mm}
\epsfig{file=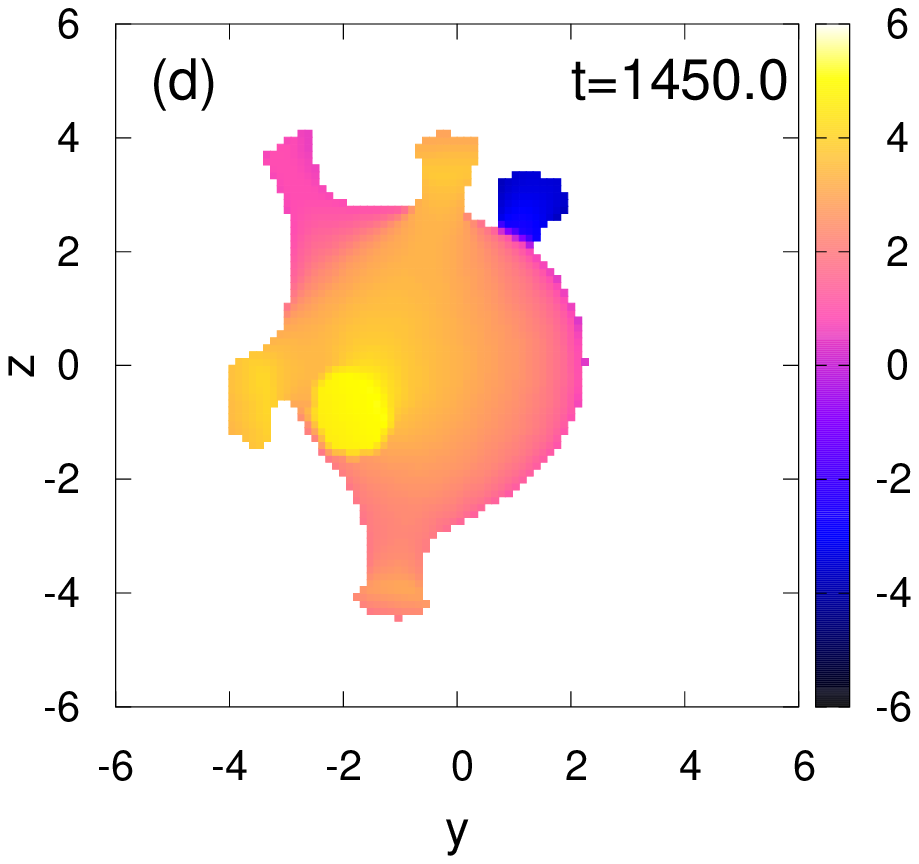, width=42mm}\\
\epsfig{file=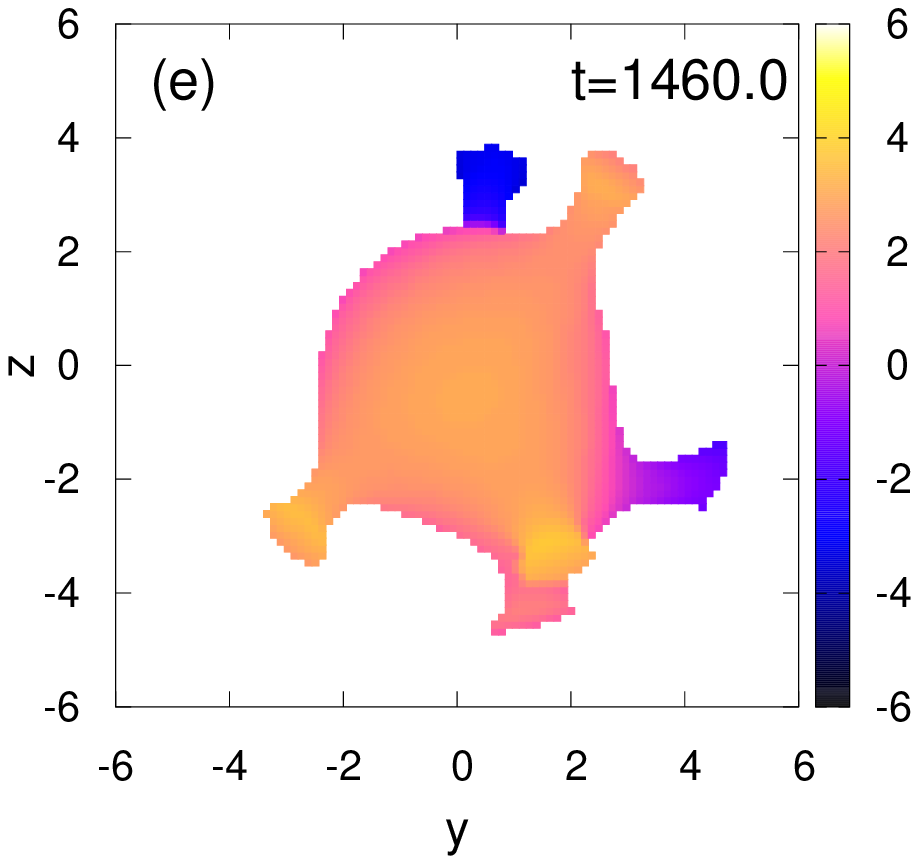, width=42mm}
\epsfig{file=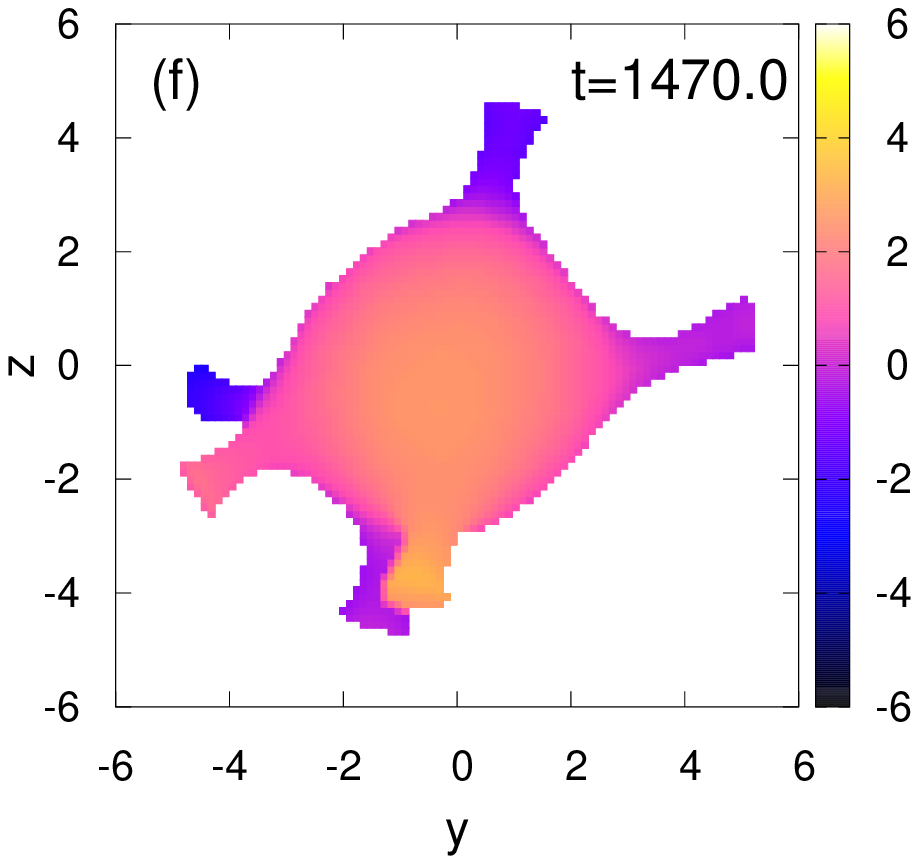, width=42mm}
\end{center}
\caption{A nonstationary bubble with three pairs of attached vortices. 
The parameters are: $\lambda=1.1$, $g_{11}=1.0$, $g_{22}=0.5$, $g_{12}=1.0$, 
$n_1=332.5$, $n_2=145.7$, and $\mu_1=18$. In this example the dynamics of vortices
is rather fast and dramatic, while the numbers of particles are relatively small.}
\label{V3_example2} 
\end{figure}

Three and more pairs of attached vortices have been observed to demonstrate, depending 
on parameters and initial conditions, both a regular and a highly dramatic and unstable
dynamics. The regular dynamics corresponds to the simplest case when a symmetric bundle 
of several vortices is in a nearly stationary rotation around $z$-axis, and also to a
slightly more nontrivial case when the rotation is around a precessing instant axis 
(see Fig.3 and video \cite{video3} as an example). Such quiet regimes are more typical 
for larger surface tension, for example with $g_{22}=0.8$. Contrary to that, with 
$g_{22}=0.5,0.6$ the bubble-vortex complexes often behave irregularly, so the mutual
arrangement of vortices evolves in a complicated and seemingly unpredictable manner. 
Using again comparison with athletic exercises, we may say that such a performance could 
be only possible for a fantastic, perfectly supple clown possessing more than two pairs 
of legs/arms. Indeed the movements look sometimes rather funny (see video \cite{video4}). 
An example of a sequence of changes for a bubble with three pairs of attached filaments 
is presented in Fig.4 (extracted from video \cite{video4}). 
This numerical experiment also demonstrates that required numbers of trapped atoms 
may be relatively small for possibility of the phenomenon under consideration. 
So, with a realistic trap length $(l_{\rm tr}/a_{11})\approx 100$, we have 
$N_1\approx 3.3\times 10^4$ and $N_2\approx 1.5\times 10^4$.

It should be said again that during unstable dynamics, the ends of two vortices 
with opposite signs often approach one another and in some cases they can connect and 
break away from the bubble, thus forming a separate filament (two examples are seen 
in videos \cite{video4,video5}). Very easy it occurs in the case of weak surface 
tension (in particular, when $g_{22}=0.3$). 
If surface tension is sufficiently strong, such behavior is typical mainly for systems 
with relatively small cores. The vortex filaments in the mantle are then rather energetic 
and in fully three-dimensional regime, so that they have opportunity to be tilted at 
large angles to the local bubble normal, thus initiating the process of separation. 

Contrary to that, a relatively large heavy core and thin mantle make the vortices 
short (while the flows of the first component approximately two-dimensional along 
the spheroid) and directed strictly perpendicular, and their dynamics is then
qualitatively similar to motion of point vortices on a spheroidal surface. 
However, the question of quantitative applicability of that analogy requires
further thorough investigation, because interactions of vortices with potential 
oscillations of the mantle may still remain important. In general, this situation 
resembles the vortex-antivortex physics in shell-shaped ordinary BECs \cite{shell-BEC}. 
Fig.4 corresponds to an intermediate regime between fully three-dimensional 
and  effectively two-dimensional regimes.

\begin{figure}[b]
\begin{center}
\epsfig{file=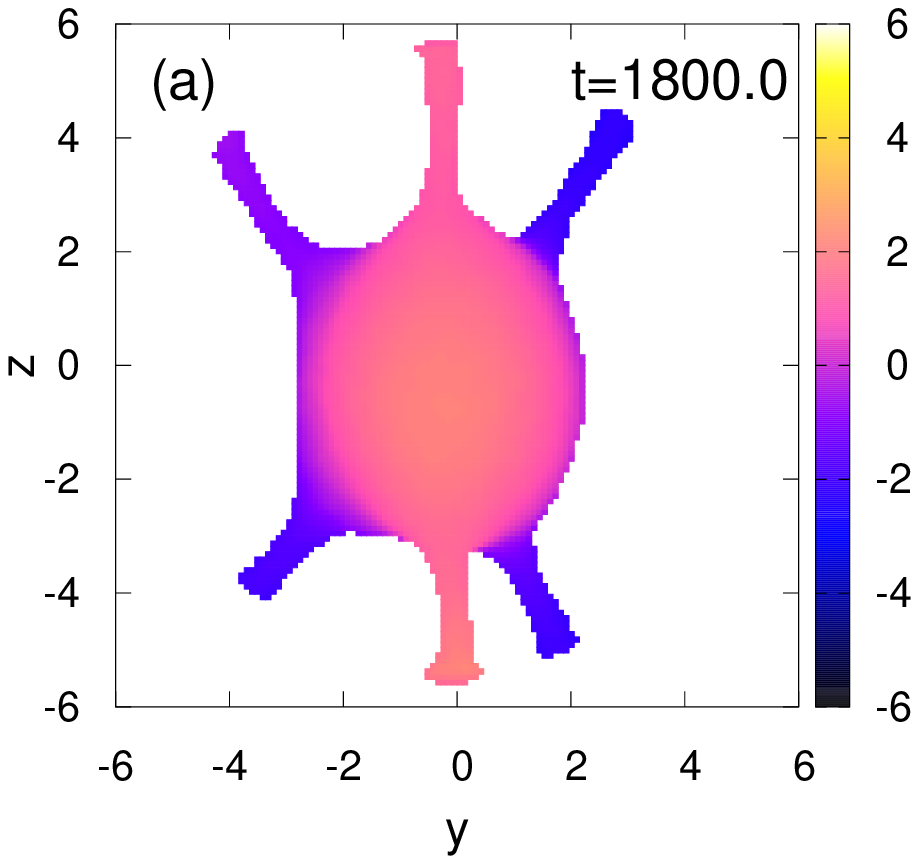, width=42mm}
\epsfig{file=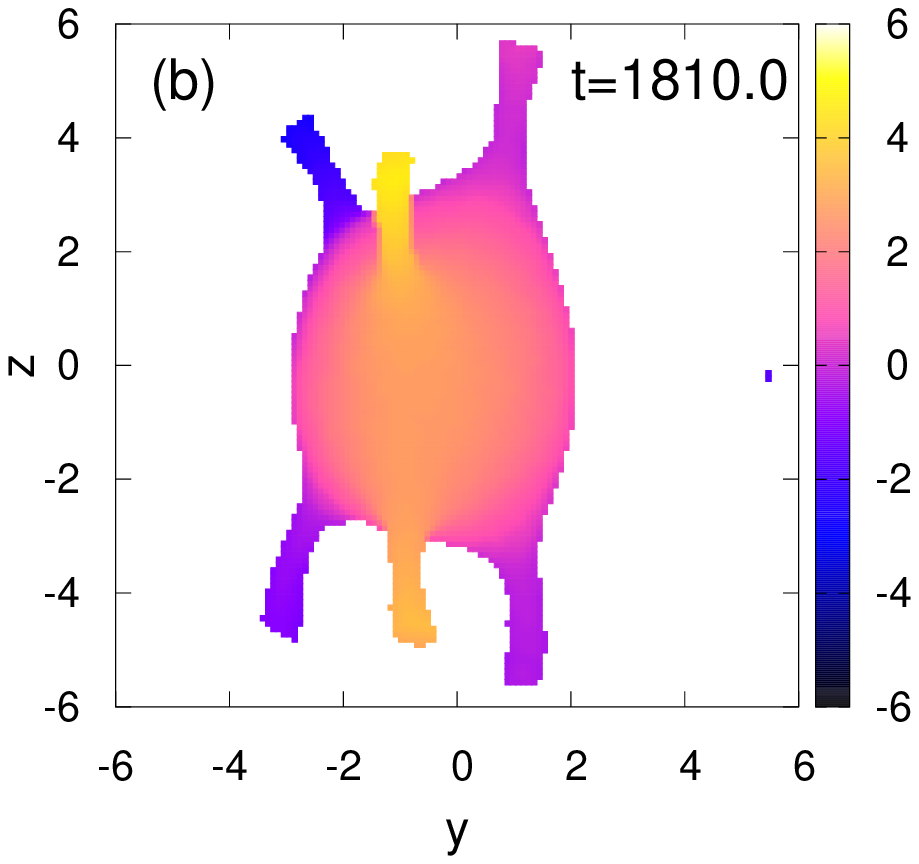, width=42mm}\\
\epsfig{file=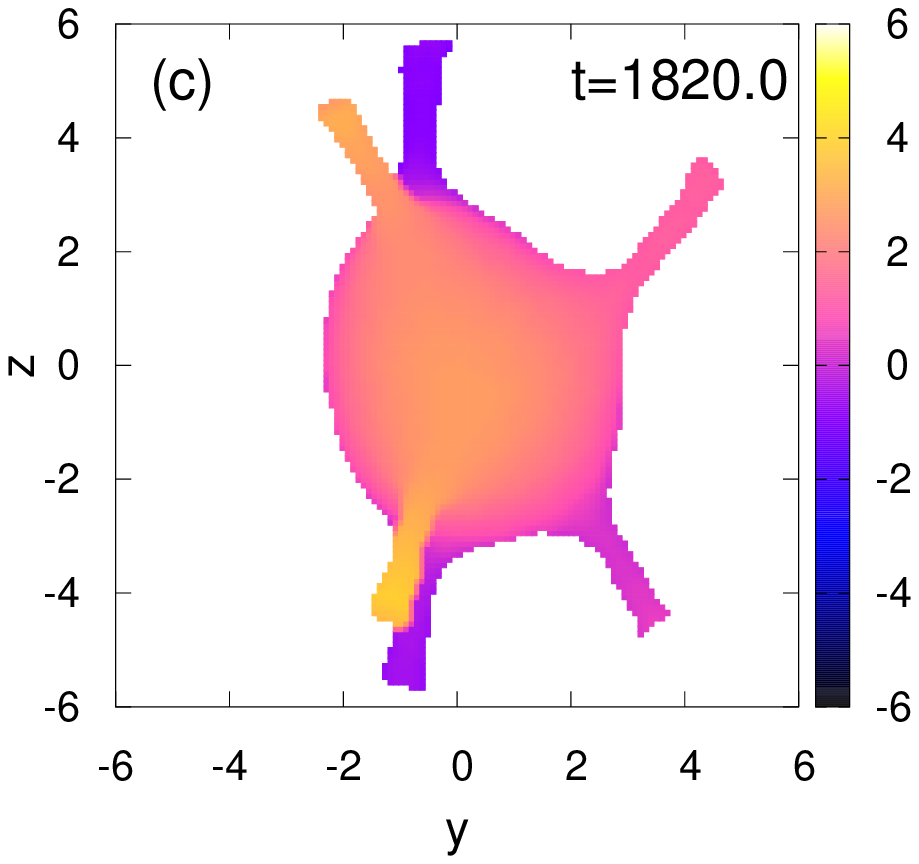, width=42mm}
\epsfig{file=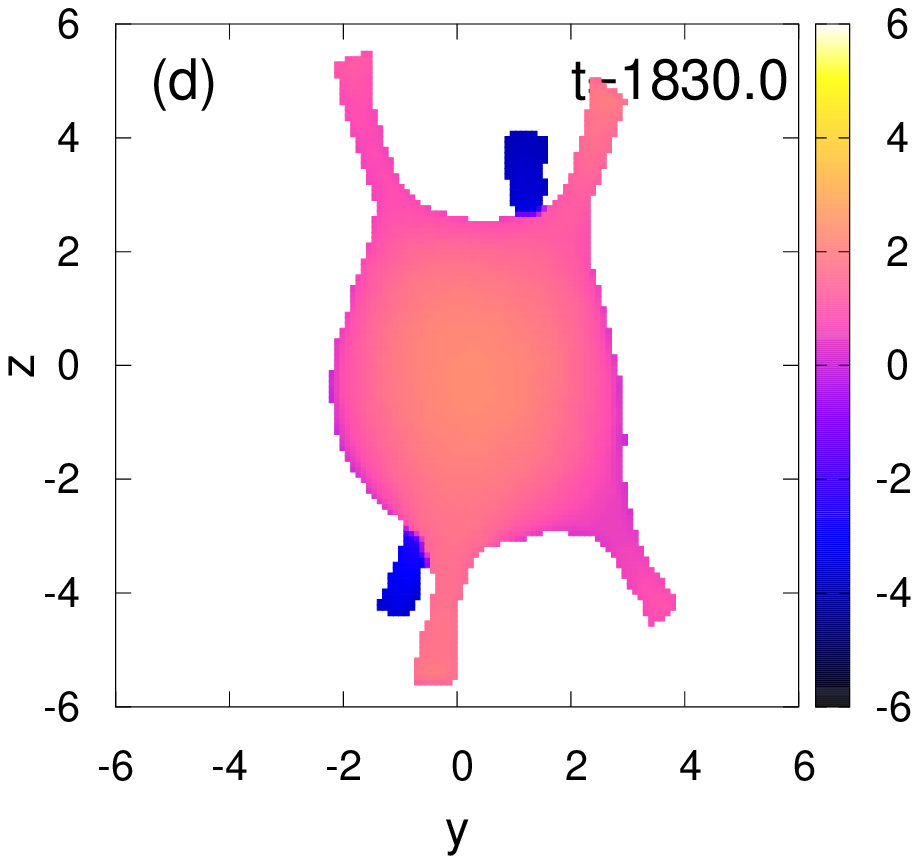, width=42mm}
\end{center}
\caption{
A bubble with three pairs of attached vortices produced by rotating trap
with the rotation angle $\phi(t)=0.08\cdot 2\pi\big(\sqrt{t^2+400^2}-400\big)$,
and the transversal anisotropy $\nu(t)=0.05 t^2/(t^2+400^2)$.  
The remaining parameters are: $\lambda=1.0$, $g_{11}=1.0$, $g_{22}=0.6$, 
$g_{12}=1.2$, $n_1=894.7$, $n_2=182.5$, and $\mu_1=24$. In this figure, the outer
ends of vortices are truncated by the condition $\rho_{1 \rm{eq}}({\bf r})>0.3$.}
\label{V3_rotating_trap} 
\end{figure}

From all the above examples it has been clear that the dynamics of attached vortices
can be approximately finite-dimensional, in the sense that a finite number of soft 
degrees of freedom is effectively involved. Apparently, the motion of attached vortices 
strongly depends on their initial arrangement. Numerically, it is easy to prepare a 
practically arbitrary initial state. However, it should be stressed here that our 
numerical initialization procedure has nothing to do with a required real experimental 
procedure to produce bubbles with attached vortices. But at this point we may recall 
that ``rotating'' trap potentials 
$$
V({\bf r},t)=(x^2+y^2+\lambda^2 z^2)/2 +\nu(t)\mbox{Re}\big[(x+iy)^2e^{-2i\phi(t)}\big]
$$
are known to ``pull'' vortex filaments from the periphery towards the condensate axis.
Here $\phi(t)$ is the rotation angle, while $\nu(t)$ is an anisotropy parameter 
in $(x,y)$ plane. Suppose we initially have a nearly equilibrium state with a
short vortex filament located near the outer boundary of the mantle. 
As the trap begins rotate, the filament approaches the core, gets attached 
to it and then forms a vortex-antivortex pair. The core remains vortex-free
in this process. At least several numerical simulations (not illustrated here) 
have demonstrated the validity of such scenario for configurations similar 
to that shown in Fig.1, including the asymmetry around $z$ axis. Angular velocity
of the trap is an important parameter in such situations. If we make the rotation faster, 
then two, three or a larger number of vortex pairs can be attached to the bubble
as the result of a complicated and essentially non-equilibrium transient process,
with each vortex being oriented at some not very large angle to $z$ axis (see an 
example in Fig.\ref{V3_rotating_trap}). 
Thus, rotating trap seems as one of the possible experimental ways. 
As to the more general problem how to produce in the laboratory a bubble with several
arbitrary arranged attached vortices, it requires a separate and deep 
investigation including efforts of highly qualified experimentalists. Therefore 
it is beyond the present purely theoretical consideration.

\section{Conclusions}

To summarize, specific wall-vortex complexes in a two-component BEC, in combination 
with its compact, stably stratified trapped configuration, have been theoretically 
introduced in this work. These are long-lived bubbles with attached vortices existing 
in wide parametric ranges and demonstrating quite interesting dynamical behavior in 
the numerical experiments. Both regular and irregular regimes have been observed.
Since we have here a new example of rich physics contained in a relatively simple 
(at least for numerical investigation) structure, further theoretical efforts in this 
direction are necessary, including development of an analytical approach.

The parameters of our simulations are quite realistic, so future experimental
implementation of such systems seems possible. A challenge for experimentalists 
will be perhaps to produce a binary condensate without small droplets and multiple 
domains, but with just a single core at the center, and after that to imprint  
several vortices into the shell. Rotating trap is a possible technique.

\end{document}